\documentclass[prl,twocolumn,aps,showpacs]{revtex4}

\usepackage{epsfig,amsmath,amssymb,amsfonts,mathrsfs,latexsym,graphicx}

\usepackage[normalem]{ulem}  
\usepackage[dvips]{color}

\begin{document}

\title{Constraints on Neutron Star Crusts From 
Oscillations in Giant Flares}

\author{Andrew W. Steiner}
\affiliation{Joint Institute for Nuclear Astrophysics, National
Superconducting Cyclotron Laboratory and the \\ Department of Physics and
Astronomy, Michigan State University, East
Lansing, MI; steinera@pa.msu.edu}

\author{Anna L. Watts}
\affiliation{Astronomical Institute ``Anton Pannekoek'', University of
  Amsterdam,
  \\ PO Box 94249, 1090 GE Amsterdam, Netherlands;
  A.L.Watts@uva.nl}

\begin{abstract}
We show that the fundamental seismic shear mode, observed as a
quasi-periodic oscillation in giant flares emitted by
highly-magnetized neutron stars, is particularly sensitive to the
nuclear physics of the crust. The identification of an oscillation at
$\approx 30$ Hz as the fundamental crustal shear mode requires a
nuclear symmetry energy that depends very weakly on density near
saturation. If the nuclear symmetry energy varies more strongly with
density, then lower frequency oscillations, previously identified as
torsional Alfv\'en modes of the fluid core, could instead be
associated with the crust. If this
 is the case, then future
observations of giant flares should detect oscillations at around 18
Hz. An accurate measurement of the neutron skin thickness of lead will
also constrain the frequencies predicted by the model.
\end{abstract}



\maketitle

Giant x-ray flares from highly magnetized neutron stars are powered by
catastrophic reconfigurations of the decaying field. Pinning of the
rapidly evolving field to the solid crust triggers an associated
starquake and generates global seismic vibrations \citep{Duncan98},
detectable as quasi-periodic oscillations (QPOs) in the
x-ray afterglow \cite{isr05, str05, wat06, str06}.

The current picture is that the QPOs, which range in frequency from 18
Hz up to 1800 Hz, result from torsional (twisting) motions of the
star. The lowest (18, 26 Hz) frequencies were initially associated
with the magnetized fluid core, while the other ($>$ 28 Hz)
frequencies were interpreted as shear modes of the solid crust. The
problem is actually more complicated, as the fluid core admits an
Alfv\'en continuum, and crust and core are coupled by the strong
magnetic field. However even in this more complex system frequencies
close to the pure crustal frequencies can still emerge, from global
modes in which crust motion dominates \citep{gla06, lee07, lev07}. If
this picture is correct, then the observed flare QPO frequencies probe
the shear properties of the neutron star crust. This is particularly
exciting because the frequencies can be measured relatively
accurately, to within a few percent or better.

In this paper, we calculate the frequency of shear oscillations of the
neutron star crust, and show that they depend sensitively on a
particular aspect of the nuclear physics input: the nuclear symmetry
energy. The symmetry energy, the energy cost of creating an isospin
asymmetry in nucleonic matter, is one of the most significant
uncertainties in the description of the
crust~\cite{Steiner05,Oyamatsu07,Steiner08}. The sensitivity of the
flare QPO frequency to the symmetry energy implies that they can
constrain the properties of the nuclear symmetry energy. We also make
a novel connection to the proposed measurement of the neutron-skin
thickness of a lead nucleus (the difference between the neutron and
proton radii), to be performed next year at Jefferson Lab in the
PREX~\cite{Horowitz01,Michaels05} experiment.

At lower densities, the outer crust of the neutron
star consists of nuclei embedded in a sea of degenerate electrons. As
density rises, nuclei become progressively heavier and more
neutron-rich. Once density increases past the neutron drip point,
$4\times 10^{11}$ g/cm$^3$, it becomes energetically favourable for
neutrons to drip out of nuclei. This begins the inner crust region of
the neutron star, where exotic nuclei are embedded in a sea of
superfluid neutrons. At a density of around $1.5\times 10^{14}$
g/cm$^3$, nuclei are no longer favored and dissolve into
their constituents. 

For the model of the inner neutron star crust, we employ the liquid
droplet model from Ref.~\cite{Steiner08}, as updated in
Ref.~\cite{Souza08}. This model consistently describes the nuclei and
the dripped neutrons in the crust with one equation of state for
homogeneous nucleonic matter, and is very similar to that employed for
the description of matter in core-collapse
supernovae~\cite{Lattimer91}. For the homogeneous nucleonic matter
EOS, we use the Skyrme model~\cite{Skyrme59}, which enables us to
describe matter over a large enough density range to obtain masses and
radii, and also can describe neutron matter at very low densities in
the inner crust.

The determination of the crust-core transition density is
difficult because the relevant energy surfaces are very flat near the
transition, so we choose to fix the transition density at 0.07 fm$^{-3}$.
Increasing this value to 0.1 fm$^{-3}$ decreases the frequency of the
radial overtones (which depend strongly on crust thickness), but
changes the fundamental crust mode frequency only by a percent or so.
We choose to use the outer crust model from~Ref.~\cite{Baym71}, but
have checked that using alternate models from Ref.~\cite{rus06} does
not change the inferred QPO frequencies (the shear mode frequencies
are largely insensitive to uncertainties in the neutron drip
point~\cite{epaps}).

The shear modulus of the crust \cite{str91} is  
 \begin{equation}
\label{shear}
\mu = \frac{0.1194}{1 + 0.595(\Gamma_0/\Gamma)^2}\frac{n_i (Ze)^2}{a}
\, .
\end{equation}
Here $Z$ is the atomic number of the ions, $n_i$ is the ion density,
and $a = (3/4\pi n_i)^{1/3}$ is the average inter-ion spacing. The
parameter $\Gamma = (Ze)^2/ak_B T$, where $T$ is the temperature,
measures the ratio of the Coulomb to thermal energies, and we use
$\Gamma=\Gamma_0 = 173$ to determine the upper boundary of the
crust~\cite{far93}. Equation (\ref{shear}), derived using a Monte
Carlo simulation of the Coulomb interactions in a neutron star crust,
assumes that the contribution to the shear modulus from
neutron-lattice interactions and neutron-neutron interactions is
small.

Neutron stars and their magnetospheres admit many types of vibration,
driven by different restoring forces. The identification of the QPOs
excited in magnetar flares with crust shear modes is based on several
factors. Shear modes have a lower excitation energy (compared to
vibrations involving bulk compression), and damp sufficiently slowly
to explain the observed QPO durations \cite{dun98}. Excitation of
crust motion is particularly likely if, as argued by
\cite{thompson95}, the yielding of a strained crust triggers the
flares. Coupling of a twisting crust to the external field also
provides a plausible mechanism for modulating the x-ray emission
\cite{str05}. Most importantly the observed QPOs match expectations
from models in terms of both frequency and the scaling between
different harmonics. Coupling of crust and core by the magnetic field
does complicate the system but, as demonstrated by \cite{lev07}, does
not prevent the emergence of the natural (uncoupled) frequencies. We
can therefore compute crust shear mode frequencies assuming free slip
between crust and core.

We employ the simple Newtonian perturbation model for torsional shear
oscillations used by \cite{pir05, wat07}. This model uses a
plane-parallel geometry with constant gravitational acceleration, $g$,
computed beforehand using the TOV equations. The Newtonian equations
of hydrostatic equilibrium then determine the crust density profile.
Using a slab rather than spherical geometry allows us to incorporate
the magnetic field without difficulty; we assume a constant field
$\mathbf{B} = B\hat{z}$. In order to correct for the error that this
introduces, and to ensure that we recover the correct spherical
geometry frequencies in the zero field limit we follow \cite{sam07}
and mimic a spherical geometry by setting $\nabla^2_\perp \xi =
-[(l+2)(l-1)/R^2] \xi$, $l$ being the standard angular quantum number.

For pure toroidal shear modes, which are incompressible and have no
vertical component of displacement, the perturbation equations for the
horizontal displacement $\xi$ reduce to

\begin{equation}
\frac{\left(\mu \xi^\prime \right)^\prime}{\rho}  + v_A^2 \xi^{\prime \prime} +
 \left[\omega^2\left(1
    + \frac{v_A^2}{c^2}\right) - \frac{(l^2+ l-2)\mu}{\rho R^2}\right] \xi
= 0 \, ,
\end{equation}
where $v_A = B/(4\pi\rho)^{1/2}$ is the Alfv\'en speed. The shear
speed is $v_s = (\mu/\rho)^{1/2}$. We assume a periodic time
dependence $\exp(i\omega t)$, $\omega$ being the frequency, and
correct for gravitational redshift to obtain observed
frequency. Primes indicate derivatives with respect to $z$.

\begin{figure}
\includegraphics[scale=0.42]{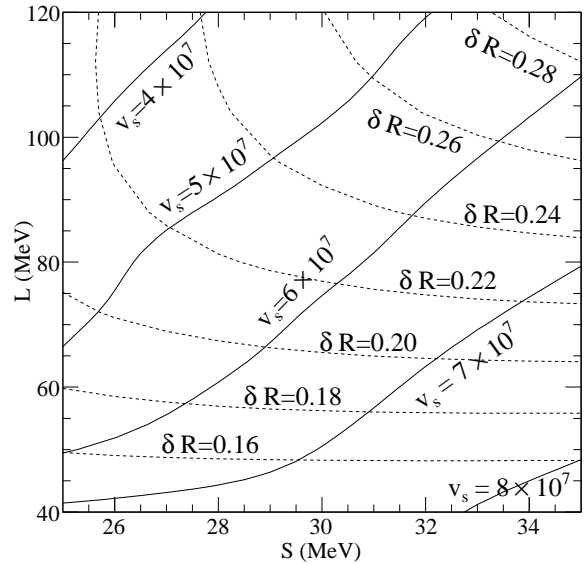}
\caption{Shear speeds $v_s$ (solid lines) and the neutron skin
  thickness of lead $\delta R$ (dashed lines) as a function of the
  symmetry energy. The abcissa is the magnitude of the symmetry energy
  at the nuclear saturation density and the ordinate is the derivative
  of the symmetry energy. }
\label{fig:0}
\end{figure}

Using the schematic model of Ref.~\cite{Steiner08}, we can construct a
neutron star crust from an arbitrarily chosen symmetry energy. This
crust model includes some corrections to the nuclear masses due to the
surrounding neutron gas. Fig.~\ref{fig:0} displays the connection to
the nuclear symmetry energy as a function of density,
$E_{\mathrm{sym}}(n)$, relating the magnitude of the symmetry energy,
S, at the nuclear saturation density, $n_0=0.16~\mathrm{fm}^{-3}$, and
a parameter related to the derivative of the symmetry energy, $L=3 n_0
\partial E_{\mathrm{sym}} / \partial n |_{n=n_0}$, to the shear speed,
$v_{s}$. The shear speed is large if the magnitude of the symmetry
energy is large and the density dependence is small. Also displayed is
the relationship to the neutron skin thickness of lead, ($\delta R$),
from the Typel-Brown correlation~\cite{Typel01}, which depends most
strongly on $L$.

\begin{figure}
\hspace*{-4mm}
\includegraphics[scale=0.35]{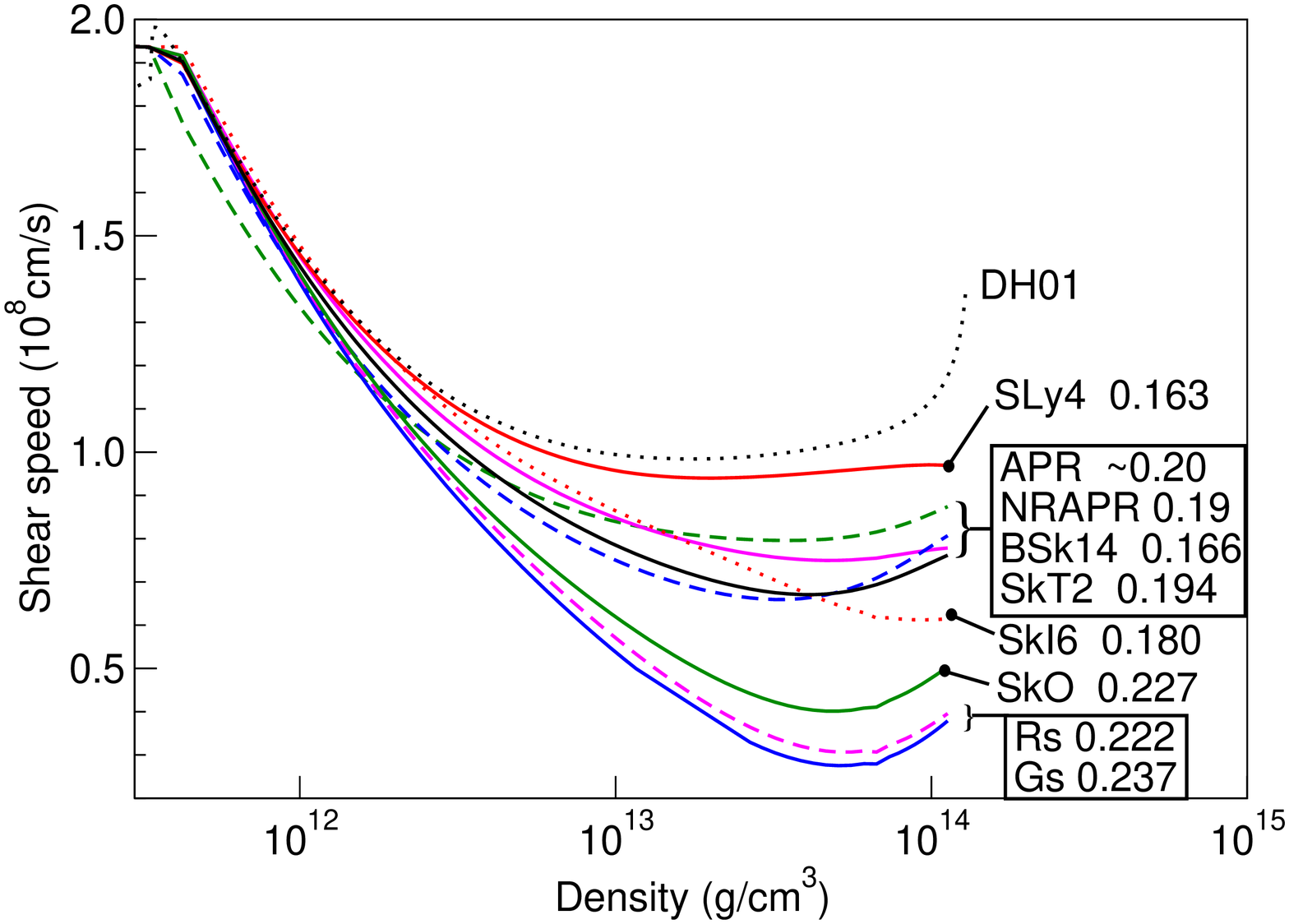}
\vspace*{-8mm}
\caption{The shear speeds as a function of density for the crust
  models used in this work, and the shear speed of Ref.~\cite{dou01}
  (labelled DH01), which
  was based on SLy4. Also shown is the neutron skin thickness of lead,
$\delta R$, for each model.}
\label{fig:1}
\end{figure}

For a full oscillation model, we need a complete model for hadronic
matter up to the densities in the center of neutron stars. The shear
speeds as a function of mass density are given in Fig.~\ref{fig:1} for
the Skyrme models, BSk14~\cite{Goriely07}, Gs~\cite{Friedrich86},
NRAPR~\cite{Steiner05}, Rs~\cite{Friedrich86}, SLy4~\cite{Chabanat95},
SkI6~\cite{Reinhard95}, SkO~\cite{Reinhard99}, SkT2~\cite{Tondeur84}.
We also use the model from APR ~\cite{Akmal98}, for which $\delta R$
has been estimated at 0.20 fm~\cite{Steiner05}. Previous calculations
such as those in Refs.~\cite{pir05,sam07} were based on the crust of
Ref.~\cite{dou01}. This also used the Skyrme model
SLy4~\cite{Chabanat95} and is shown as a dotted line, labelled DH01.
The difference between that result and our model, derived from the
same Skyrme interaction, provides an estimate of the uncertainties in
the construction of the masses of the neutron-rich nuclei in the
crust, given the same underlying nucleon-nucleon interaction. That
difference is smaller than the variation arising from the
nucleon-nucleon interaction itself, which is as much as a factor of 4.
This large variation in the shear speed is the source of the
sensitivity of the QPO frequencies to the nuclear physics input. As
expected from Fig.~\ref{fig:0}, there is a correlation between the
neutron skin thickness in lead, with smaller skins generally giving
larger shear speeds. The upcoming PREX measurement will help constrain
the frequencies predicted by the model, reducing the uncertainty
originating from our lack of knowledge of the symmetry energy.

\begin{figure}
\hspace*{-3mm}
\includegraphics[scale=0.34,clip]{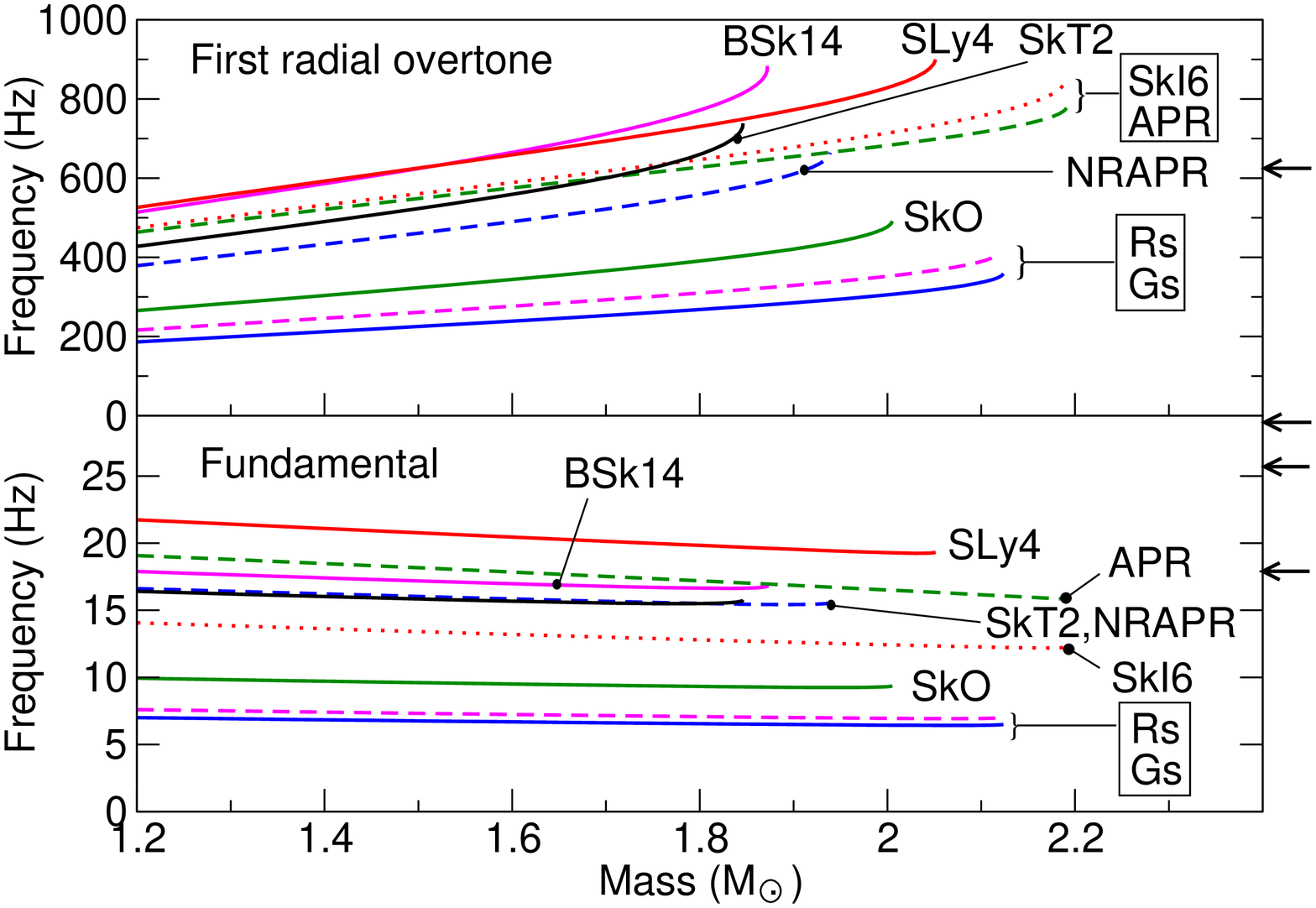}
\vspace*{-8mm}
\caption{The crust oscillation frequencies as a function of neutron
  star mass, for both the fundamental ($n=0,l=2$) torsional shear mode
  and the first radial ($n=1$) overtone. The curves end at the maximum
  mass. The arrows on the right indicate QPO frequencies measured during
  the 2004 hyperflare from SGR 1806-20 \cite{isr05, wat06, str06}.}
\label{fig:2}
\end{figure}

Figure~\ref{fig:2} shows the frequencies of the fundamental $n=0, l=2$
crust shear mode and the first ($n=1$) radial overtone as a function
of neutron star mass for the various crust models. Also shown in this
figure are some of the QPO frequencies measured during the 2004
hyperflare from SGR 1806-20. The dramatic effect of uncertainty in the
symmetry energy on shear speed is evident in the spread of mode
frequencies. The frequency of the fundamental mode is larger for
smaller neutron skin thicknesses, however all of the models have
difficulty explaining a fundamental above 22 Hz. Previous studies
including \cite{pir05}, which interpreted the 28 Hz QPO in SGR 1900+14
and the 29 Hz QPO in SGR 1806-20 as the fundamental, employed the
higher shear speed model of \cite{dou01}, thereby yielding higher
frequencies. Once the correct $l$ scaling is taken into account,
however, even this crust model would necessitate a rather low stellar
mass \cite{sam07}. The models that we have considered, which have a
symmetry energy that varies more strongly with density, cannot explain
a fundamental frequency at 28-29 Hz. The only observed QPO that would
fall into the predicted range is the 18 Hz QPO detected in the SGR
1806-20 hyperflare \cite{isr05, wat06} and previously interpreted as a
torsional mode of the magnetized fluid core \cite{isr05}. 

Relativistic perturbation calculations predict very similar
frequencies for the fundamental to those generated by Newtonian
calculations \cite{sam07}. The mode frequencies are also largely
insensitive to magnetic field effects, since the electron Fermi
momentum in the inner crust is much larger than the energy spacing
between Landau levels except at fields larger than the $10^{14-15}$ G
implied for magnetars. Note also that the fundamental frequency is
nearly mass-independent, which means that uncertainties in the neutron
star mass will not spoil the connection between the frequency and the
nuclear symmetry energy. It is not well known how finite-size
effects and the presence of extremely-deformed nuclei (i.e. nuclear
pasta) may affect the shear modulus. Recent calculations by
\cite{Horowitz08b}, which re-examine Equation (\ref{shear}) would
exacerbate the problem since they indicate that shear modulus is
actually lower by $\sim 10$\%. This would reduce mode frequencies by a
further 3-4\%. Finally, the structure of the neutron-rich nuclei
are not well understood and this also adds a systematic uncertainty
to our results.

Our crust model thus suggests that a revision of the previous
interpretation of the 28 Hz mode as the fundamental shear mode may be
necessary. One alternative is that the fundamental mode may have a
lower frequency (an idea also explored in \cite{sot07}). For this to
be plausible, it should be possible to fit the sequence of QPOs
detected in each flare with the expected scaling in $l$. In the SGR
1806-20 giant flare, a QPO was detected at 18 Hz. If this is the
fundamental ($l=2$), then a reasonable fit to the higher frequencies
can be obtained with the following mode identifications: 29 Hz
($l=3$), 92 Hz ($l=10$), 150 Hz ($l=16$). The one QPO that would not
fit the sequence is the 26 Hz QPO, but this could also be accommodated
if the fundamental were even lower, at $\approx 11$ Hz. Such low
values are predicted by some of our models, although it might then be
difficult to explain the 625 Hz QPO as an $n=1$ radial overtone.
Alternatively, the 26 Hz QPO is a core-dominated mode.

No lower frequency QPOs were detected in the SGR 1900+14 hyperflare.
However the sequence of detected QPOs would be compatible with a
fundamental at $\approx 18$ Hz with the following mode
identifications: 28 Hz ($l=3$), 53.5 Hz ($l=6$), 84 Hz ($l=9$) and 155
Hz ($l=17$). As for SGR 1806-20, a lower frequency fundamental at
$\approx 11$ Hz could also be accommodated. With this in mind, we
performed a thorough rotational phase and energy-dependent search of
the data from the {\it Rossi X ray Timing Explorer} for the SGR
1900+14 hyperflare to search for lower frequency QPOs. No detections
were made, but due to poorer data quality we could not search for QPOs
as weak as those detected in SGR 1806-20. At 18 Hz, for example, we
were able to set a 3 sigma upper limit on a QPO amplitude of 7\% rms:
in the SGR 1806-20 hyperflare the 18 Hz QPO had an amplitude of 4\%.

It is interesting that an 18 (or 11) Hz fundamental frequency might be
able to explain both mode sequences. If this is a common feature then
we would expect to detect such a feature in future giant flare
lightcurves. The lowering of crust shear mode frequencies and their
proximity to the expected frequencies of magnetic torsional modes does
however complicate efforts to use magnetar QPOs to measure interior
field strengths \cite{isr05, sot07}. A sequence of modes, including
overtones, is likely to be necessary to enable correct mode
identification. There are several other possible effects on the shear
modulus, including that of shell effects, frozen
impurities~\cite{jon99, jon01}, superfluidity and entrainment
\citep{andersson09}, and the magnetic field which we have not
addressed. A more definitive mode assignment will have to wait until
these effects are better understood, or more data is obtained.

AWS was supported by JINA at MSU under NSF-PFC grant PHY 08-22648, and
by NASA ATFP grant NNX08AG76G. ALW thanks JINA for hospitality. We
also thank Ed Brown, Lars Samuelsson and participants of the
University of Washington Institute for Nuclear Theory workshop ``The
Neutron Star Crust and Surface: Observations and Models'' for
stimulating discussions.

\bibliographystyle{apsrev}
\bibliography{crustphys}

\end{document}